# RADIOMICS-GUIDED MULTIMODAL SELF-ATTENTION NETWORK FOR PREDICTING PATHOLOGICAL COMPLETE RESPONSE IN BREAST MRI

*Jonghun Kim*[1,2]        *Hyunjin Park*[†1,2]

[1] Department of Electrical and Computer Engineering, Sungkyunkwan University, Suwon, Korea
[2] Center for Neuroscience Imaging Research, Institute for Basic Science, Suwon, Korea

## ABSTRACT

Breast cancer is the most prevalent cancer among women and predicting pathologic complete response (pCR) after anti-cancer treatment is crucial for patient prognosis and treatment customization. Deep learning has shown promise in medical imaging diagnosis, particularly when utilizing multiple imaging modalities to enhance accuracy. This study presents a model that predicts pCR in breast cancer patients using dynamic contrast-enhanced (DCE) magnetic resonance imaging (MRI) and apparent diffusion coefficient (ADC) maps. Radiomics features are established hand-crafted features of the tumor region and thus could be useful in medical image analysis. Our approach extracts features from both DCE MRI and ADC using an encoder with a self-attention mechanism, leveraging radiomics to guide feature extraction from tumor-related regions. Our experimental results demonstrate the superior performance of our model in predicting pCR compared to other baseline methods.

***Index Terms*—**multimodality, radiomics, attention mechanism, pathological complete response, breast MRI

## 1. INTRODUCTION

Breast cancer ranks as the most common cancer among female patients [1]. Pathologic complete response (pCR) refers to a condition where, after pre-operative anti-cancer treatment, subsequent pathological examination reveals the complete disappearance of the original cancer cells [2]. Predicting pCR holds a significant correlation with the long-term prognosis of patients. Specifically, patients who achieve pCR are at a reduced risk of recurrence or mortality [2, 3]. Furthermore, if pCR can be accurately predicted, it enables the customization of therapeutic strategies for individual patients. For instance, patients who are anticipated to have a high probability of achieving pCR might benefit from reduced intensities of chemotherapy or might be candidates for alternative treatment modalities [4].

Deep learning has been gaining significant attention in the medical field in recent years. Particularly in medical imaging-based diagnosis, it has demonstrated remarkable efficacy and accuracy. Given that each imaging modality possesses its unique characteristics and advantages, it is crucial to leverage multimodality imaging in deep learning-based diagnoses to enhance the precision and reduce the limitations of the diagnostic process [5].

†Corresponding author: Hyunjin Park (email: hyunjinp@skku.edu )

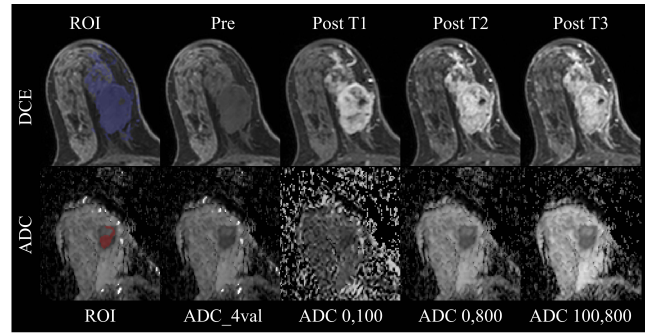

**Fig. 1.** Description of DCE MRI and ADC map for breast cancer. Colored regions are the tumor regions from two different patients. In DCE (top), 'pre' represents the image taken before contrast agent injection. Post T1, T2, and T3 depict images captured at successive time points after the contrast injection. In ADC (bottom), the ADC_4val represents the ADC computed from four DWI b-values: 0, 100, 600, and 800. The other ADC maps represent the ADC calculated from two specific b-values of DWI.

Dynamic contrast-enhanced (DCE) imaging involves the acquisition of magnetic resonance imaging (MRI) at various delayed time intervals after the administration of a contrast agent. It is instrumental in visualizing and quantifying the presence and distribution of proliferative blood vessels in cancerous tissue [6]. **Fig. 1** (top) illustrates DCE images captured at regular intervals post-contrast injection.

The Apparent Diffusion Coefficient (ADC) map, derived from diffusion-weighted imaging (DWI) MRI using multiple b-values, quantitatively represents the degree of water molecule diffusion within tissues. Since cancerous tissues possess high cell density, the diffusion of water molecules within them is restricted. ADC values reflect this diffusion limitation. Thus, regions with low ADC values are highly suggestive of cancerous tissue [7]. **Fig. 1** (bottom) depicts the ADC derived from DWIs of various b-values. Consequently, DCE and ADC, each with their unique advantages and characteristics, when employed together in breast cancer diagnosis, can offer more precise and comprehensive information.

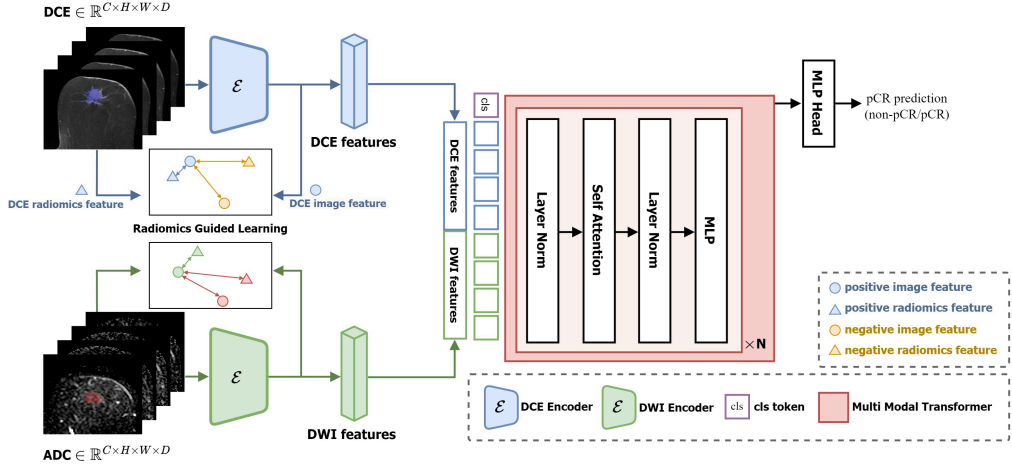

**Fig. 2.** Illustration of proposed radiomics guided multimodal self-attention network.

In this study, we propose a model to predict pCR using DCE MRI and the ADC map. Our contributions are as follows:

First, we extract features from both DCE and ADC using an encoder and further employ a self-attention mechanism, facilitating the model to learn the relationships between features from these two modalities.

Second, by extracting radiomics features [8,9] from the tumor region of interest (ROI), we ensure that the encoder effectively detects the tumor and its surrounding regions in the MRI, thereby guiding the model to extract features pertinent to pCR.

Third, we visually confirm that the image features, guided by radiomics, accurately focus on the vicinity of the tumor. Through experimental validation, we demonstrate that our model exhibits superior performance in predicting pCR.

## 2. METHOD

In our methodology for predicting pCR, we utilize DCE and ADC in a multimodal fashion. For the DCE, we stack one pre-contrast image along with three post-contrast images across the channel direction. Moreover, our ADC maps utilize four different ADC maps, each derived from combinations of DWI images with varying b-values, arranged in the channel direction for our analysis. Hence, both ADC and DCE can be represented as three-dimensional volume images, each composed of four channels, formalized as $\mathbf{I}_{DCE}, \mathbf{I}_{ADC} \in \mathbb{R}^{C \times H \times W \times D}$.

### 2.1. Multimodal self-attention network

To execute pCR prediction in a multimodal breast MRI, we propose a Radiomics-guided Multimodal self-Attention network (RaMA-net) as illustrated in **Fig. 2**. We first extract image features using a 3D CNN Encoder. This encoder operates with radiomics-guided learning, which is explained in Section 2.2. The output from the encoder can be represented as:

$$\mathbf{X} = \mathcal{E}(\mathbf{I}) \quad (1)$$

where $\mathcal{E}$ is 3D CNN encoder, image $\mathbf{I} \in \mathbb{R}^{C \times H \times W \times D}$ and feature $\mathbf{X} \in \mathbb{R}^{C_{\mathcal{E}} \times h \times w \times d}$ with $h = \frac{H}{8}, w = \frac{W}{8}$ and $d = \frac{D}{8}$.

Subsequently, we aggregate the features along the channel direction at each voxel location and feed them into a transformer. At this point, the feature $X$ is reshaped from $X \in \mathbb{R}^{C_{\mathcal{E}} \times h \times w \times d}$ to $X \in \mathbb{R}^{N \times C_{\mathcal{E}}}$ where $N = h \times w \times d$. Subsequently, the features are concatenated and a class token (cls) is added to the start of the sequence for effective self-attention. The class token is denoted as $z_{cls}$ with dimensions $\mathbb{R}^{d_{dim}}$ where $d_{dim}$ represents the embedding dimension. The sequence, **Z**, can be described as follows:

$$\mathbf{Z} = [z_{cls}, T_{DCE}^1, T_{DCE}^2, \ldots, T_{DCE}^N, T_{ADC}^1, T_{ADC}^2, \ldots, T_{ADC}^N] \quad (2)$$

The tokens $T_{DCE}^1, \ldots, T_{DCE}^N$ are derived by unfolding the DCE feature $X_{DCE}$, while $T_{ADC}^1, \ldots, T_{ADC}^N$ are tokens obtained from the ADC feature $X_{ADC}$. The sequence **Z** obtained from the embedding process is subsequently processed through a sequence of transformer layers. Every layer in this sequence integrates multi-head self-attention (**MSA**) [10], layer normalization (**LN**), and **MLP** blocks in the given order:

$$\tilde{\mathbf{Z}}_k = \mathbf{MSA}(\mathbf{LN}(\mathbf{Z}_k)) + \mathbf{Z}_k \quad (3)$$

$$\mathbf{Z}_{k+1} = \mathbf{MLP}(\mathbf{LN}(\tilde{\mathbf{Z}}_k)) + \tilde{\mathbf{Z}}_k \quad (4)$$

The **MLP** layer consists of two linear projections with a Gaussian error linear unit (GELU) activation function in between. This method allows the model to capture the related information between DCE and ADC features, while simultaneously encapsulating a summarized representation of the entire sequence within the class token.

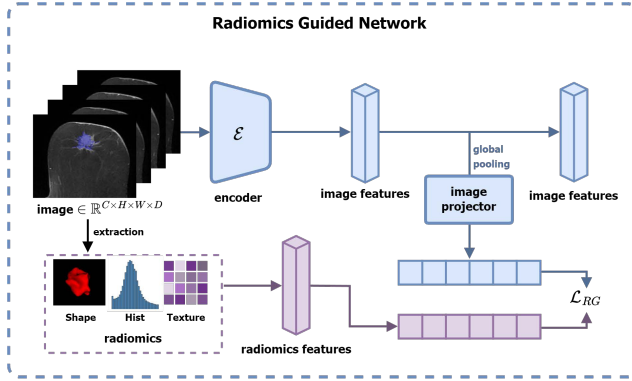

Fig. 3. Illustration of the proposed Radiomics Guided Network (RGN). The radiomics features include shape, histogram, and texture features from the ROI. Using these radiomics features, the encoder is trained by contrastive learning with image feature.

## 2.2. Radiomics-guided learning

We guide our encoder to learn representations related to tumors using radiomics information, enabling it to better predict pCR. This learning process is facilitated through contrastive learning [11, 12] as depicted in **Fig. 3**. Features from the same subject's image and radiomics are treated as positive pairs, while features from different subjects are treated as negative pairs. The process can be represented as:

$$\mathcal{L}_{RG} = -\log \frac{\exp(sim(v_i, v_r)/t)}{\sum_{k=1, k\neq i}^{2B} \exp(sim(v_i, v_r)/t)} \quad (5)$$

where $sim$ denotes the cosine similarity, $t$ denotes the measurement on a normalized temperature scale. The vector $v_i$ is derived from the encoder after global average pooling and subsequently passing through the image projector. In contrast, $v_r$ is a vector derived from the radiomics feature. The vectors $v_i$ and $v_r$ from the same subject form a positive pair, while image feature vectors and radiomics feature vectors from different batches are considered negative pairs.

## 2.3. Interpretability module

To verify whether the image encoder has effectively learned representations near the tumor guided by radiomics features, we employ a patch-based interpretability module, similar to methods used in previous studies [13, 14], to visualize the similarity between radiomics features and image features. When the patches of the image feature pass through global pooling and the image vector is guided by the radiomics feature, we can calculate the similarity value for each patch by comparing its image vector to the radiomics feature vector. This process is illustrated in **Fig. 4**. Each patch's image vector, after passing through the image projector, calculates its similarity with the radiomics feature. By overlaying this heatmap onto the MRI for visualization, we can verify if the representation around the tumor has been effectively learned with the guidance of radiomics.

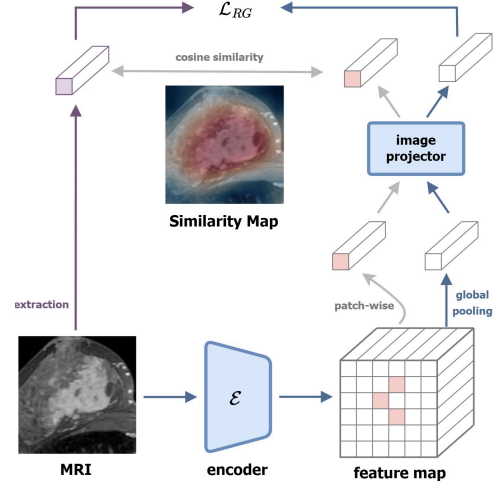

Fig. 4. Illustration of the operation of the interpretability module. The cosine similarity is computed between each patch of the image feature map and the radiomics feature. This resulting heatmap is then overlaid on the MRI for visualization.

## 3. EXPERIMENTS

### 3.1. Dataset

For our model's training and evaluation, we used the pre-treatment time point before neoadjuvant chemotherapy from the ACRIN-6698 dataset [15-17]. The dataset comprises 191 breast cancer patients (59 pCR, 132 non-pCR). The DCE images come with a 512×512 matrix size, matrix size, in-plane resolution upto 1.4 mm, slice thickness upto 2.5 mm, and phase duration of 80-100 seconds ensuring at least 8 minutes of continuous post-injection acquisition. We utilized one pre-contrast image along with three post-contrast images. These post-contrast images are used at intervals of two phase durations, resulting in a set of four images, which are concatenated channel-wise for training. The ADC images have a matrix size of 256×256 with an in-plane resolution ranging from 1.7 to 2.8 mm and a slice thickness between 4 to 5 mm. The ADC map is constructed from a combination of DWI images with four distinct b-values (b = 0,100,600, and 800). The specific combinations are outlined in **Fig. 1**.

### 3.2. Implementation Details

Both DCE and ADC images are cropped and resized focusing on the tumor, resulting in a consistent size of 128×128×64 pixels for preprocessing. All comparison models were set with a batch size of 2. Every experiment was conducted using 5-fold cross-validation, with the dataset divided into training and validation sets per patient. The Adam [18] optimizer was employed with $\beta_1$ set at 0.9 and $\beta_2$ at 0.999, and a learning rate of 0.0001. The model was built using PyTorch[1] version 1.12.1, and MONAI[2]. All experiments were conducted on 4 TITAN XP GPUs, each with 12GB of memory.

---

1. https://pytorch.org/
2. https://monai.io/

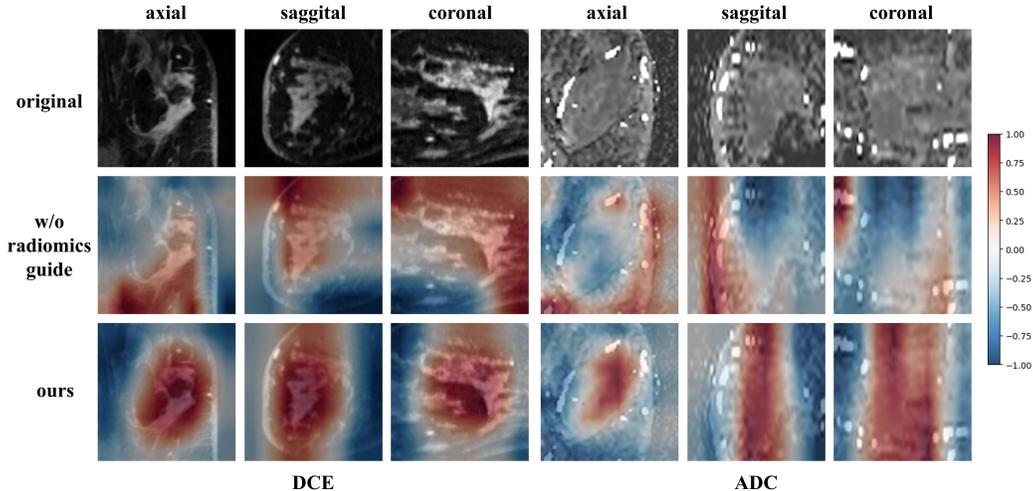

**Fig. 5.** Visualization of the similarity heatmap between radiomics features and image features. Since it is a 3D image, it displays overlays of the MRI and heatmap in three views: axial, sagittal, and coronal. Left: DCE. Right: ADC.

**Table. 1.** The performance of the comparison models when using only one modality. Models with the best performance in each modality are highlighted in bold. The symbol ± represents the standard deviation from 5-fold cross-validation.

| Model | Modality | Accuracy | AUC |
|---|---|---|---|
| DenseNet121 | DCE | 0.7401±0.017 | 0.6961±0.015 |
| DN + Rad | DCE | 0.7457±0.015 | 0.7218±0.012 |
| **RGN (ours)** | DCE | **0.7752±0.016** | **0.7226±0.014** |
| DenseNet121 | ADC | 0.7405±0.021 | 0.6584±0.025 |
| DN + Rad | ADC | 0.7205±0.019 | 0.6574±0.022 |
| **RGN (ours)** | ADC | **0.7457±0.020** | **0.6730±0.021** |

**Table. 2.** Performance comparison in an ablation study when training using DCE and ADC as a multi-modality approach.

| Encoder | Radiomics | Transformer | Accuracy | AUC |
|---|---|---|---|---|
| DenseNet121 | ✗ | ✗ | 0.7735±0.018 | 0.7107±0.017 |
| | ✓ | ✗ | 0.7862±0.015 | 0.7315±0.018 |
| | ✗ | ✓ | 0.7950±0.016 | 0.7479±0.017 |
| | ✓ | ✓ | **0.8181±0.017** | **0.7642±0.019** |

### 3.3. Results

We employed DenseNet121 [19], a light weight model, as our 3D image encoder to efficiently extract 3D image features. Additionally, to demonstrate the effectiveness of the radiomics guidance, we introduced a model that concatenates radiomics features with image features similar to a previous study [20] and compared it for classification. The performance of these comparison models, trained on a single modality, is detailed in **Table 1**. When imaging features were concatenated with radiomics features, there was a noticeable improvement in performance. However, it is worth noting that our approach led to a significant boost in performance.

### 3.3. Ablation Study

We conducted additional experiments to assess the importance of the radiomics guidance and transformer when training the model using multimodality input. The performance comparison from this ablation study is detailed in **Table 2**. In this experiment, we performed classification by concatenating both image features when not using the Transformer. Through these experiments, we noted an improvement in performance with DCE and ADC as multimodality input, and a dramatic enhancement when adding radiomics guidance and Transformer structure.

### 3.4. Visualization of interpretability

We visualized whether the encoder effectively learned important tumor vicinity and key texture representations for pCR prediction by utilizing the interpretability module. **Fig. 5** illustrates this visualization. When examining the heatmap, it becomes evident that without receiving the radiomics guide, there is a lack of noticeable similarity between image features and radiomics features in the vicinity of the tumor at various patch locations. However, our proposed approach effectively integrates the radiomics guide, resulting in a higher similarity between image features and radiomics features, indicating successful learning of representations around the tumor.

### 4. CONCLUSION

We have proposed a radiomics-guided multimodal self-attention network for classifying breast cancer patients' pCR predictions by utilizing DCE and ADC. Additionally, to capture essential tumor-related representations for pCR prediction, we introduced the radiomics guidance network, which exhibited significant performance improvements. While our results indicate that performance can be enhanced by simply concatenating image features when leveraging multimodality input in training, we have demonstrated that learning complementary relationships through attention mechanisms can lead to the acquisition of meaningful information and substantial performance gains.

## 5. ACKNOWLEDGMENTS

This study was supported by National Research Foundation (NRF-2020M3E5D2A01084892), Institute for Basic Science (IBS-R015-D1), AI Graduate School Support Program (2019-0-00421), ICT Creative Consilience program (IITP-2023-2020-0-01821), and the Artificial Intelligence Innovation Hub program (2021-0-02068).

## 6. REFERENCES

[1] Lei, Shaoyuan, et al. "Global patterns of breast cancer incidence and mortality: A population-based cancer registry data analysis from 2000 to 2020." Cancer Communications 41.11 (2021): 1183-1194.

[2] Spring, Laura M., et al. "Pathologic complete response after neoadjuvant chemotherapy and impact on breast cancer recurrence and survival: a comprehensive meta-analysis." Clinical cancer research 26.12 (2020): 2838-2848.

[3] Huang, Min, et al. "Association of pathologic complete response with long-term survival outcomes in triple-negative breast cancer: a meta-analysis." Cancer Research 80.24 (2020): 5427-5434.

[4] Freitas, Ana Julia Aguiar de, et al. "Molecular biomarkers predict pathological complete response of neoadjuvant chemotherapy in breast cancer patients." Cancers 13.21 (2021): 5477.

[5] Li, Rongjian, et al. "Deep learning based imaging data completion for improved brain disease diagnosis." Medical Image Computing and Computer-Assisted Intervention–MICCAI 2014: 17th International Conference, Boston, MA, USA, September 14-18, 2014, Proceedings, Part III 17. Springer International Publishing, 2014.

[6] Padhani, Anwar R. "Dynamic contrast-enhanced MRI in clinical oncology: current status and future directions." Journal of Magnetic Resonance Imaging: An Official Journal of the International Society for Magnetic Resonance in Medicine 16.4 (2002): 407-422.

[7] Le Bihan, Denis. "Apparent diffusion coefficient and beyond: what diffusion MR imaging can tell us about tissue structure." Radiology 268.2 (2013): 318-322.

[8] Gillies, Robert J., Paul E. Kinahan, and Hedvig Hricak. "Radiomics: images are more than pictures, they are data." Radiology 278.2 (2016): 563-577.

[9] Lambin, Philippe, et al. "Radiomics: the bridge between medical imaging and personalized medicine." Nature reviews Clinical oncology 14.12 (2017): 749-762.

[10] Vaswani, Ashish, et al. "Attention is all you need." Advances in neural information processing systems 30 (2017).

[11] Chen, Ting, et al. "A simple framework for contrastive learning of visual representations." International conference on machine learning. PMLR, 2020.

[12] Park, Taesung, et al. "Contrastive learning for unpaired image-to-image translation." Computer Vision–ECCV 2020: 16th European Conference, Glasgow, UK, August 23–28, 2020, Proceedings, Part IX 16. Springer International Publishing, 2020.

[13] Turgut, Özgün, et al. "Unlocking the Diagnostic Potential of ECG through Knowledge Transfer from Cardiac MRI." arXiv preprint arXiv:2308.05764 (2023).

[14] Boecking, Benedikt, et al. "Making the most of text semantics to improve biomedical vision–language processing." European conference on computer vision. Cham: Springer Nature Switzerland, 2022.

[15] Newitt, David C., et al. ACRIN 6698/I-SPY2 Breast DWI [Data set]. The Cancer Imaging Archive. (2021) https://doi.org/10.7937/TCIA.KK02-6D95

[16] Newitt, David C., et al. "Test–retest repeatability and reproducibility of ADC measures by breast DWI: Results from the ACRIN 6698 trial." Journal of Magnetic Resonance Imaging 49.6 (2019): 1617-1628.

[17] Partridge, Savannah C., et al. "Diffusion-weighted MRI findings predict pathologic response in neoadjuvant treatment of breast cancer: the ACRIN 6698 multicenter trial." Radiology 289.3 (2018): 618-627.

[18] Kingma, Diederik P., and Jimmy Ba. "Adam: A method for stochastic optimization." arXiv preprint arXiv:1412.6980 (2014).

[19] Huang, Gao, et al. "Densely connected convolutional networks." Proceedings of the IEEE conference on computer vision and pattern recognition. 2017.

[20] Joo, Sunghoon, et al. "Multimodal deep learning models for the prediction of pathologic response to neoadjuvant chemotherapy in breast cancer." Scientific reports 11.1 (2021): 18800.